# On Physical Web Browser

Dmitry Namiot
Lomonosov Moscow State University
Moscow, Russia
dnamiot@gmail.com

Manfred Sneps-Sneppe
Ventspils University College
Ventspils, Latvia
manfreds.sneps@gmail.com

*Abstract*—In this paper, we present the Physical Web Browser project for web applications depending on the environment. At this moment, many of users all over the world visit websites using their mobile devices only. Any mobile device (e.g., smartphone) has sensors to capture the environmental information. This information (context) could be analyzed and used within the web applications. There are, at least, two models for using this information. Firstly, we can use context information for data gathering requests in a web application. In this case, the output depends on the context. It is the classical model for context-aware data retrieval. In the second model, we can use context-aware data for improving user experience (for changing and tuning user interfaces). In other words, we present a way for the adaptation of web applications depending on the environment.

## I. INTRODUCTION

There are two paradigms related to our job. The first one is so-called context-aware computing. In the original paper, introduced the term 'context-aware', the author described context as location, identities of nearby people and objects, and changes to those objects [1]. So, technically, context is any measurable data that can be used to characterize the situation of an entity. And an entity here is a place, person, or object that is considered relevant to the interaction between a user and an application [2]. Note, that this definition includes the user and applications themselves. This description makes it easier for an application developer to enumerate the context for a given application scenario [3].

The second paradigm is so-called Ambient Intelligence (AMI) [4]. AMI is a paradigm which it aims multidisciplinary development physical environments where different electronic objects intelligently respond to the presence of people [5]. AMI targets the creation of sensitive, adaptive electronic environments that respond to the actions of persons and objects and cater for their needs. AMI approach includes the entire environment (all physical objects) and links (associates) it with human interaction. As a result, we can expect an extended and more intuitive interaction, enhanced efficiency, increased creativity, etc. In our case, the interaction targets mobile web applications.

In general, AMI encourages the creation of intelligent environments for embedded systems. These systems use sensors embedded to capture information from the real environment. In our case, we have sensors embedded into mobile devices. Context and context-awareness are central issues to AMI. The availability of context and the use of context in interactive applications offer new possibilities to customize (adopt) applications and systems "on-the-fly" to the current situation [6]. Context information can influence and change interactive systems. Authors in [7, 8] introduce the conception of implicit human computer interaction. It is the interaction of a human with the environment and with artifacts, which is aimed to accomplish a goal [8]. Within this interaction, the system acquires implicit input from the user and may present implicit output back to the user. The implicit input here includes actions and behavior of humans. The word "implicit" highlights the fact that data are not primarily regarded as interaction with a computer, but captured, recognized and interpret by a computer system as input [8]. So, sensors on a mobile device are a perfect example of implicit input. Similar to this, an implicit output is an output of an application that is not directly related to an explicit input and which is seamlessly integrated with the environment and the task of the user. The basic idea of the implicit input is that the system can perceive the user's interaction with the physical environment and also the overall situation in which an action takes place [8].

There are many examples of AMI in smart homes and smart cities [9], telemedicine [10], transport [11, 12], living environments [13, 14] and education [15].

In our research, we would like to investigate how to deal with environment information in mobile web applications. The reasons are very transparent. For native applications, it is not a problem to include sensor-based measurements into processing. In the same time, web access gives the flexibility: we do not need to develop an application, upload it to the store (marketplace) and ask potential users to install it.

The rest of the paper is organized as follows. In Section II, we describe related works. In Section III, we discuss web access to sensor data, present the Physical Web Browser project and existing prototype.

## II. RELATED WORKS

Technically, we can highlight the following areas for context-aware and ambient intelligence applications.

Smart agents. Also, it could be named as triggers or proactive applications [16, 17]. It is the natural approach to use events or more general situations to trigger the start (stop) of applications. The common requirement is a direct connection between the context and the application that is executed. The typical actions are starting or stopping some service or raising a warning. The classical example is a control system, which carry out a predefined action when the certain context is recognized. For web applications, such action could be a push notification, for example. All proximity-based notifications systems in retail belong to this category. Also, there is certain connection to smart-spaces area [18, 19]

We can use context information as a parameter for proactive applications. The behavior of the application is then changed according to context. The simplest example of this type of application is a navigation system. The context-based parameters are the speed and current position.

As the next area for context-aware services deployment, we can mention an adaptive user interface. Actually, it is the most natural area for web applications. Having information on the current situation available, it becomes possible to build user interfaces that adapt to context [20].

With mobile devices, it is possible to take into account physical changes in the environment (e.g., light, noise). The most obvious example is device orientation (portrait or landscape). Of course, we should take into account the context of use at design time. All the assumptions about usage scenarios are made in the design process [21, 22].

By our opinion, we should pay attention to such important element as user interruption. The modern mobile applications use push notifications, for example, in network proximity tasks. For example, it is Apple's iBeacon model [23]. For push notifications, the only requirement is the preinstalled application with allowed push notifications mode. A mobile user even does not need to run this application. It means that push notification could be pushed anytime as soon as the application is still installed on the device. There is no need in the confirmation from the mobile user. The service can send notifications (and so, interrupt the user) anytime and without limitations. In this sense, the browsing looks more user-friendly. Al least during the browsing the user has the direct intention to obtain some data. It is more easily to achieve stability in the user interface without confusing the user. In the case of browsing, the proactive behavior of the system is understandable and predictable by the user even if the details are hidden.

Context information can help to enhance remote communication between people. As an example, we can mention network proximity based systems [24]. Spotique is a mobile mashup combines passive monitoring for smart phones and cloud-based messaging for mobile operational systems. Passive monitoring can determine the location of mobile subscribers (mobile phones, actually) without the active participation of the users. Mobile users do not need to mark their location on social networks (check-in).

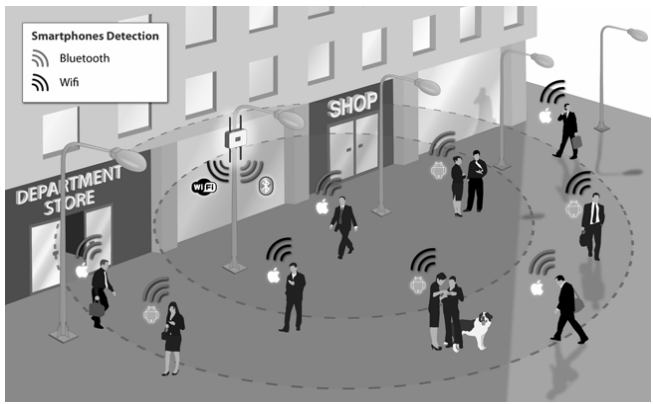

Fig.1. Smartphones detection [25]

They do not need to run on their phones the location track applications too. Cloud Messaging lets data providers (e.g., local businesses) directly deliver their information to mobile users nearby a selected point. It uses commercial out-of-the-shelf component (Meshlium Xtreme) for passive Wi-Fi monitoring (Fig. 1). The basic idea here is to filter communications according to context. And location-aware systems play the main role here [26].

Resource Management (using resources dependent on the context) is a classical application for mobile devices. The most obvious example is network switching for mobile devices (e.g., from GPRS to Wi-Fi). In the most cases, this resource management is also linked with the proximity. Some authors highlight the fact that using resources that are physically close or in a proximity of the user is central to this type of applications. The concept of physical proximity and the usage of physical space as the main criteria for ordering items (accessing them) is a very natural concept for humans [6].

The next possible are for context-aware data is enriching requests. We can talk here about any form of meta-data adding to a user request. In this connection, for example, W3C GeoLocation API is also a form of meta-data, developers can use in web applications. As another example we can mention, for example, context-aware QR-code reader [27]. It is a typical mashup - context-aware QR-code scanner is based on the modified version of open sourced scanner Zxing (Zebra Crossing). QR code presents two-dimensional (2D) barcode. In general, 2D barcodes encode some text. But in many cases, that text can represent many different things. In the most often usage, 2D barcodes encode text that represents some URL, like "http://some_domain.com/". Technically, it is a special use case for text representation, where URL's pattern is recognized by the software. QR-code reader can recognize an URL and open it in the browser. Context-aware QR-code reader keeps the basic process as is, but just adds some parameters on the final stage. In other words, the customized QR-code scanner will replace encoded value

http://some_domain.com/

with

http://some_domain.com?list_of_parameters

And this list of parameters will describe our context. For example, let us see the QR-code deployment for some indoor retail application. QR-codes will let mobile visitors download an appropriate coupon. Technically, we can prepare a separate QR-code for the each existing department (for the each existing discount program). But this process could be costly and difficult to maintain. Context-aware QR-code reader can automatically add context information (e.g., location or proximity info) to any encoded URL. It lets us use the same generic QR-code across all our installations. The URL in QR-code points to some CGI-script which can proceed HTTP GET parameters in the request and respond accordingly.

According to many researchers, the fundamental problem is how to incorporate context data into applications [28, 29]. In general, "context pull" means that the consumer of context, e.g. the user or the application) directly requests data (e.g., with the incorporated context info). It means that the consumer controls what is requested and is used. In general, it is a preferable approach for web browsing. By the way, we can pull data and use obtained information later. It not necessary to immediately refresh data even for the mobile applications. The "context push" model describes a mode where the context producing entity delivers context to the consumers. The decision when to supply the context is up to the provider. The most obvious example is push notification in mobile OS. For the context provider, this makes distributing context straightforward. As soon as a new context is available, it could be pushed to consumers. The push-model has the advantage for the context consumer does not actively have to query for context. In the pull model, consumers need to request new data periodically, even if a new context info is actually not available yet. But the biggest disadvantage is user interruption. However in terms of implementation, this can be also an enormous. As an intermediate mode, some authors propose a publish/subscribe model. A context provider can publish new data as soon as they are available, a subscriber (consumer) can read them by the own initiative. In this case, we can simply let consumers know that new data are available, without the pushing them without user's intention to obtain.

Note, that for mobile applications nowadays push approach is definitely the prevailing model. For example, the above-mentioned iBeacons use push notifications. The pull model is used for web applications. In the same time, we must mention here push notification in HTML5 [30]. Technically, in HTML5, Web Sockets and SSE (Server Sent Events) are both capable of pushing data to browsers. Web Sockets connections can both send data to the browser and receive data from the browser. A good example of an application that could use Web Sockets is a chat application. SSE connections can only push data to the browser. For example, Twitters updating timeline or RSS feed are good examples of an application that could benefit from SSE. SSEs are sent over traditional HTTP. That means they do not require a special protocol or server implementation to get working.

The Push API by W3C [31] enables sending of a push message to a web application via a push service. An application server can send a push message at any time, even when a web application or user agent is inactive. The push service ensures reliable and efficient delivery to the user agent. Push messages are delivered to a Service Worker that runs in the origin of the web application, which can use the information in the message to update local state or display a notification to the user. It is illustrated in Fig. 2.

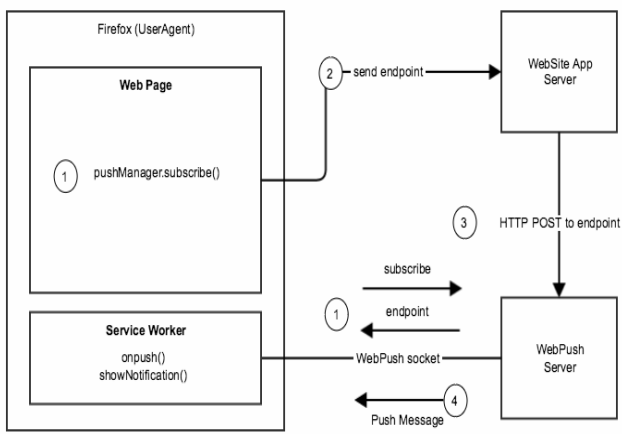

Fig. 2 Push notifications [32]

By our opinion, web technologies should be especially attractive for context-aware applications due to flexibility. We simply cannot expect to have a separate application for everything. And the browser is an ideal candidate for the commonly used access tool. In the same time, when a user accesses a Web application using the smartphone, that user may be affected by the restrictions on the hardware device itself. Environmental factors can affect the experience while the user is using the web application. They include noise, too much light, lack of light, inability to provide focus attention on small letters while a user is walking or moving, etc. We can conclude that the mobile web applications are always the same, but the environment could be changed. In this connection, some of the researchers propose so-called Web Adaptation [33, 34]. The main idea here is to generate a derived version of an original web application, capable of improving user experience in some way. Originally, it is started as mobile versions of web sites. In the most cases, this adaptation was based on form-factor based usability for mobile devices. With evolving HTML5 standards, it becomes possible to take into account geo-location, orientation, as well as accelerometer and gyroscope. Still, most of the adaptations do not depend on the environment for the mobile device, which can be variable and can negatively affect the user experience.

In this paper, we present a model of the context-aware mobile web browser. This browser should use data from sensors on a mobile phone for capturing information about the environment which can be used to web applications customization. Of course, this customization should target all the above-mentioned aspects: content, presentation and user interfaces.

### III. WEB ACCESS TO SENSORS

Technically, as a generic model for access to sensor data from web applications (web pages) we can see W3C geo-location. It means, that mobile browser should follow to the same ideology for the reuse of information about existing sensors as geo-coding in HTML5 [35]. Let us see the top-level details [36].

A function from browser's interface

*navigator.geolocation.getCurrentPosition()*

accepts as a parameter some user-defined callback (another function). The callback should be called as soon as geo-location is completed. Obtained data should be passed as parameters. Note, that the whole process is asynchronous.

In our prototype, we target network proximity applications. Network proximity is based on the detection of wireless nodes (Wi-Fi access points, Bluetooth nodes, Bluetooth Low energy tags). By the analogue with the above-mentioned model, a mobile browser can add a new interface function. E.g.,

*getNetworks()*

this function will accept a user-defined callback for accumulating network information (current network fingerprint). A good candidate for the data model is JSON. The browser will pass fingerprint as a JSON array to a user-defined callback. Each element from this array describes one network and contains the following information:

SSID - name for access point

MAC - MAC-address

RSSI - signal strength

Note, that scanning networks is an asynchronous process in mobile OS. So, callback pattern is a good fit for this. Firefox OS is closest in ideology to this approach [36]. Also, Firefox OS provides Bluetooth API [37]. It has got the similar ideology, but there is no general approach (denominator). E.g., even fields for objects are different). It should be possible, of course, to create some unified wrapper (shell), which will give a general list of networks. The biggest problem (we are not mentioning here the "popularity" for Firefox OS) is the status for both APIs. Wi-Fi API has just been scheduled yet. At the same time, the Bluetooth API exists, but it is declared preferred (privileged). Privileged APIs can be used by the operating system only. So, it could not be used in applications. The reason for this solution is security. API combines both network scanning and network connection (data exchange). It is the wrong design by our opinion. APIs functionality should be separated. The network proximity approach is not about the connectivity. Mobile OS should use two separate APIs: one for scanning (networks poll) and one for connecting. Polling for networks does not require data exchange. So, scanning API is safe, and it should not be privileged. It is simple - we should have *WiFiManager* interface (as is, and it could be privileged),

and WiFiScan with only one function getNetworks():

```
<script>
function callback function(json data ) { ... }
WiFiScan.getNetworks(callback function);
</script>
```

The callback function can loop over an array of existing networks IDs and show (hide) HTML div blocks with data related (associated) to the existing (visible) networks.

Actually, it is a fundamental question. Traditionally, wireless networks on mobile phones are used as networks. But they are sensors too. The fact that some network node is reachable (visible) is a separate issue. And it could be used in mobile applications even without the ability to connect to that node. It is the main idea behind the network proximity approach, and it is the feature (option) we suggest to embed into mobile browsers.

The obtained context information (in this use case it is the network proximity info) could be used in mobile web application via production rules (if-then blocks):

```
IF (network_fingerprint) THEN
    // show data and/or
    // perform HTTP (Ajax) requests
ENDIF
```

How can we present our rules for network proximity? As per our suggestion, each data snipped should be presented as a separate div block in HTML code. E.g., it is some like this:

```
<div id="BIG_MALL">
    // show content for this mall
</div>
```

We can use CSS styles to hide/show this block. And this CSS visibility attribute depends on the visibility of Wi-Fi (Bluetooth) nodes. Of course, CSS visibility could be changed in JavaScript. So, our rules could be implemented in JavaScript code. We can directly present the predicates in our code, or describe their parts in CSS too. HTML5 custom attributes are good candidates for new set of context-related attributes [38]. It means also, that adding some set of rules to existing web page looks like as adding (including) some JavaScript code (JavaScript file). In general, this approach could change the paradigm of designing mobile web sites. It eliminates the demand to make separate versions for local sites or events. It is enough to have one common site with local data (events, etc.) placed in hidden blocks. For network proximity, hidden blocks will be visible to mobile users in the proximity of some network nodes.

Local blocks visibility depends on the network nodes visibility and so, it depends on the current location of mobile users. E.g., for the above-mentioned example, mobile users opened BIG_MALL site being physically present in the proximity of BIG_MALL, will see different (additional) data compared with any regular mobile visitor. And CSS blocks will be able to perform different tasks: invoke requests for getting new data as well as change user interfaces. Of course, single data source (just one web site) support simplifies (makes it cheaper) the maintenance during life time.

As an one interesting approach, originally followed to this architecture, we can mention Web Intents [39]. The Web Intents formation was a client framework (everything is executed in the browser) for the monitoring (polling) and building services interaction within the application. Interactions include data exchange and transfer of control. Intents are in the core architecture of Android OS [40], but unfortunately, Web Intents development is stopped after some initial interesting experiments from Google. We should mention in this context a similar (by its conception) initiative from Mozilla Labs - Web Activities [41]. But the further status of this initiative is also unclear.

The next possible toolbox is seriously underrated in our opinion. It is a local web server. The first implementation, as far as we know, refers to the Nokia [42]. In our opinion, this is one of the most promising areas for communicating with phone sensors. A local web server is a mobile application. This application can collect sensors data and make them available for web pages in the local mobile browser. Web pages should simply incorporate (via <script> tag) some JavaScript file from the local web server. And this JavaScript file (actually, CGI script) will return JSON data from phone's sensor. The same approach could be reproduced with some ideas from the old WAP (Wireless Access Protocol). In the case of WAP, a mobile device used some intermediate server (WAP Gateway) for access to internet resources. This intermediate server (new gateway) should be able to collect sensing information from a mobile device (including wireless networks sensors). It could be a potential idea for 5G networks. The always-on mobile device saves own sensing data on a gateway, where web access (through the same gateway) could use accumulated sensing data. The collecting mobile sensors data on servers is a well-developed topic in scientific papers [43]

Authors in [44] propose a new set of XML tags for access to sensors data. They have considered a design decision is to deploy an agent reflective based on models. Such an agent is able to analyze real-time data captured by the sensors. Depending on the history and current values, the agent is able to determine if there is a change in a value of any of the states. Internally, the mobile browser Web uses an intelligent model-based reflective agent which is responsible for analyzing the captured data. The agent will process information and reports on environmental changes. The processor evaluates adaptation conditions (XML tags) included in the website. And an adaptation engine introduces changes to the Web code captured by the sensors. Here is a list of some sensing analysis functions for the agent:

a) func_moving_user: uses the latest GPS and measures accelerometer to determine whether the user is moving or not.

b) func_noise_polution: uses the latest measures microphone to determine the current level of noise.

c) func_accelerometer_trend: uses the latest measures accelerometer to determine if the device is on a stable surface.

c) func_orientation_trend: uses the latest measures orientation sensor to determine if the user is rotating the device.

d) func_pedestrian-vehicle: uses the latest measures GPS and accelerometer to determine whether the user is using a vehicle and walking.

e) func_front_camera_detection: uses the OpenCV library to determine what activity there in front of the camera.

f) func_ambient_light: uses the latest sensor values ambient light to determine the light level.

For HTML markup, they use a new set of XML tags. E.g.:

```
<Ami_adaptation
environment = "user_movement_type VEHICLE ==">

<Input id = "n" type = "text" x-webkit-speech="x-webkit-speech"
onwebkitspeechchange =
"This.form.submit ();" />
</ Ami_adaptation>
```

A new tag *Ami_adaptation* marks an updatable block. In our approach, we use custom HTML5 attributes. As a prototype, we propose a custom *WebView* for Android. On Android platform, it is possible to access from JavaScript to Java code for a web page, loaded into *WebView* control. For the network proximity, Java code will provide a list of nearby network nodes (calculate the network fingerprint). So, as an "agent" we will use a custom Java code for the selected group of sensors. The key moment here is the need for an asynchronous call from JavaScript because scanning for wireless networks (as well as access to the most of other sensors) in Java is the asynchronous process.

Let us describe this approach a bit more detailed. On Android side we activate JavaScript interface:

```
public void onCreate(Bundle savedInstanceState) {
  super.onCreate(savedInstanceState);
  WebView webView = new WebView(this);
  setContentView(webView);
  WebSettings settings = webView.getSettings();
  settings.setJavaScriptEnabled(true);
  webView.addJavascriptInterface(new MyJavascriptInterface(), "Network"); }
```

Now we can describe our Java code for getting network fingerprint. As a parameter, we will pass a name for callback function in JavaScript.

```
@JavascriptInterface
public void getNetworks(final String callbackFunction)
{ }
```

We skip the code for network scanning and demonstrate the final part only. As soon as a fingerprint in obtained, we can present it as JSON array and invoke our callback:

```
webView.loadUrl("javascript:" + callbackFunction + "('" + data + "')");
```

And on our web page, we can describe our callback function and call Java code:

```
function f_callback(json) { }
Network.getNetworks("f_callback");
```

This approach lets us obtain the network proximity info right in JavaScript (in other words, right on the web page). Actually, by the similar manner, we can work with other sensors too. It is so-called Data Program Interface [21]. Probably, we will see something similar as a standard feature in the upcoming versions of Android SDK.

IV. CONCLUSION

The paper discusses the usage of mobile sensors data right within web pages. We discuss the related works in this area and propose several approaches to the implementation of a mobile browser that can handle sensing data on the dynamic web pages. On the network proximity example, we demonstrated the practical way for incorporation sensors data into custom WebView control on Android platform. This approach could be summarized as a set of Java proxy modules for various sensors accessible right from JavaScript. As HTML integration, we propose the usage of custom HTML5 attributes.

ACKNOWLEDGMENT

We would like to thank prof. V.Vishnevsky for the valuable discussions.